\documentclass[useAMS,usenatbib,fleqn]{mn2e}

\usepackage{graphicx,color,amsmath,amssymb,amsfonts}

\newcommand{\be}{\begin{equation}}
\newcommand{\ee}{\end{equation}}
\newcommand{\bea}{\begin{eqnarray}}
\newcommand{\eea}{\end{eqnarray}}

\newcommand{\dco}{\delta_{\rm cs}}

\newcommand{\om}{\Omega_m}
\newcommand{\ob}{\Omega_b}
\newcommand{\odm}{\Omega_c}

\newcommand{\ho}{H_0}

\newcommand{\fnl}{{f_\mathrm{NL}}}
\newcommand{\fng}{{f_\mathrm{NL}^{\rm GR}}}

\newcommand{\mb}{\mathcal{Q}}

\title{Einstein's legacy in galaxy surveys}
\author[S. Camera, R. Maartens \& M.G. Santos]{Stefano Camera,$^{1,2}$\thanks{E-mail: stefano.camera@manchester.ac.uk.} Roy Maartens$^{3,4}$ \& M\'ario G. Santos$^{3,5,2}$\\
$^1$Jodrell Bank Centre for Astrophysics, School of Physics \& Astronomy, The University of Manchester, Manchester M13 9PL, UK\\
$^2$CENTRA, Instituto Superior T\'ecnico, Universidade de Lisboa, 1049-001 Lisboa, Portugal\\
$^3$Physics Department, University of the Western Cape, Cape Town 7535, South Africa\\
$^4$Institute of Cosmology \& Gravitation, University of Portsmouth,  Portsmouth PO1 3FX, UK\\
$^5$Square Kilometre Array SA, Cape Town 7405, South Africa}
\date{}

\begin{document}

\date{Accepted 0000 --- 00. Received 0000 --- 00; in original form 0000 --- 00}

\pagerange{\pageref{firstpage}--\pageref{lastpage}} \pubyear{2014}

\maketitle

\label{firstpage}
\begin{abstract}
Non-Gaussianity in the primordial fluctuations that seeded structure formation  produces a signal in the galaxy power spectrum on very large scales. This signal contains vital information about the primordial Universe, but it is very challenging to extract, because of cosmic variance and large-scale systematics---especially after the \textit{Planck} experiment has already ruled out a large amplitude for the signal. Whilst cosmic variance and experimental systematics can be alleviated by the multi-tracer method, we here address another systematic---introduced by not using the correct relativistic analysis of the power spectrum on very large scales. In order to reduce the errors on $\fnl$, we need to include measurements on the largest possible scales. Failure to include the relativistic effects on these scales can introduce significant bias in the best-fit value of $\fnl$ from future galaxy surveys.
\end{abstract}

\begin{keywords}
cosmology: large-scale structure of the universe---early Universe---cosmological parameters---observations---radio lines: galaxies---relativistic processes.
\end{keywords}

\section{Introduction}

One of the most important open questions in cosmology is whether or not the primordial fluctuations are Gaussian. Primordial non-Gaussianity (PNG) imprints a characteristic feature, via the bias $b$, in the galaxy power spectrum $P_g=b^2P$.  This feature is a growth of power $\propto \fnl k^{-2}$ on large scales. The excess power is `frozen' on super-Hubble scales during the evolution of the galaxy overdensity, and is unaffected by nonlinearity on small scales. 

The best current constraints on PNG are from cosmic microwave background (CMB) temperature and polarisation measurements by the \textit{Planck} satellite \citep[][]{Ade:2015ava}. For the local form of PNG, which has the strongest impact on galaxy bias, 
 \be\label{fpg}
 \fnl = 1.0\pm 6.5,
\ee
i.e.\ $\sigma(\fnl)_{\rm \textit{Planck}}=6.5$. Here we use the large-scale structure convention, $f_{\rm NL}^{\rm(LSS)}\simeq 1.3 f_{\rm NL}^{\rm(CMB)}$ \citep{Camera:2014bwa}.
The \textit{Planck} constraint rules out inflationary models with large PNG. In order to discriminate amongst the remaining models we need to significantly reduce the error $\sigma(\fnl)$. 
 
Galaxy surveys are not yet competitive with \textit{Planck}. Future surveys covering a large fraction of the sky and reaching high redshifts, such as  \textit{Euclid}\footnote{\texttt{www.euclid-ec.org}} \citep{Laureijs:2011gra,Amendola:2012ys} and the Square Kilometre Array\footnote{\texttt{https://www.skatelescope.org}} (SKA) \citep[][]{Dewdney:2009,Maartens:2015mra},
will be able to probe many more modes than the CMB. 
The future of PNG constraints lies with huge-volume surveys of the large-scale cosmic structure \citep{Camera:2013kpa,Camera:2014dia}, provided that systematics can be controlled and approximations in the  modelling of bias and haloes can be improved.
Such surveys will be able to access horizon-scale modes, thus exploiting the growth of the PNG signal on these scales. 

In addition to experimental systematics, there is also a potential theoretical systematic that arises when general relativistic (GR) effects on large scales are ignored in the data analysis. This theoretical systematic can be avoided by using an accurate analysis that includes all known effects \citep{Camera:2014bwa}.
What is the origin of these GR effects? The answer is described below, but briefly it is as follows.

First, there is a nonlinear GR correction to the primordial Poisson equation that requires a correction to the observed $\fnl$: 
\be
f_{\rm NL}^{\rm obs} = \fnl+\fng \simeq \fnl -2.2\,. \label{fcor}
\ee
In particular, for the simplest single-field models, with $\fnl\simeq0$, the signal in the galaxy power spectrum would be $f_{\rm NL}^{\rm obs}\simeq-2.2$.

Secondly, there are GR corrections to the standard linear power spectrum arising from observing on the past lightcone. The observed galaxy number counts contain not only the well-known Kaiser redshift-space distortions, but also further relativistic contributions from lensing convergence, Doppler terms, Sach-Wolfe (SW) and integrated SW (ISW) terms and a time-delay term. On sub-Hubble scales, the redshift-space distortions and lensing can make significant contributions, while the other terms are typically negligible. However on scales near and beyond the Hubble horizon $H^{-1}(z)$, the other GR terms can become important. 

When $\fnl$ is not large, as indicated by \eqref{fpg}, galaxy surveys need to cover huge volumes in order to detect the tiny primordial signal. In this paper, we show that for future surveys, the theoretical analysis must be accurate enough to correctly identify any primordial signal. Our focus is not on forecasting for particular experiments. Instead, we use a reference survey to analyse  {\em the bias on the best-fit value of $\fnl$ due to neglect of GR effects}.  Our results indicate that it is essential to include all GR effects in order not to bias the determination of $\fnl$.

\section{PNG with relativistic effects} 

We parametrise the local-type deviation from Gaussianity in the primordial curvature perturbation via 
\begin{equation}\label{eqn:ng}
\Phi=\varphi+\fnl\left(\varphi^2-\langle\varphi^2\rangle\right),
\end{equation}
where $\varphi$ is a first-order Gaussian perturbation. Local PNG induces a large-scale modulation of the small-scale formation of haloes in the cold dark matter, producing a scale (and redshift) dependence in the halo bias. On large scales (roughly, beyond the equality scale), this leads to the substitution \citep{Dalal:2007cu,Matarrese:2008nc,Giannantonio:2011ya}
\begin{equation}
b(z)\to b(z)+\Delta b(z,k),
\end{equation}
where
\begin{equation}
\Delta b(z,k)=[b(z)-1]\frac{3\om\ho^2q\delta_{\rm cr}}{k^2T(k)D(z)}\,\fnl. \label{eq:bias-NG}
\end{equation}
The factor $q=O(1)$ reflects residual uncertainty in modelling the PNG modification to the halo mass function, and we follow \citet{Giannantonio:2011ya} in setting $q=1$ in the absence of more accurate modelling. This does not affect our main result about the bias introduced when neglecting GR effects.    $\om=\ob+\odm$ is the total matter fraction at $z=0$,  $\delta_{\rm cr}\simeq1.69$ is the critical matter density contrast for spherical collapse, $T(k)$ is the matter transfer function  and $D(z)$ is the linear growth function of density perturbations, normalised to $D(0)=1$. 
The key $k^{-2}$ term  in \eqref{eq:bias-NG} comes from relating the matter overdensity $\delta$ to $\Phi$ via the Poisson equation. 

In the standard approach to constraining PNG via the galaxy bias, we use a sub-Hubble (`Newtonian') analysis and the Kaiser approximation to the redshift space distortions (RSD),
\be\label{dk}
\delta_g^z = (b+\Delta b)\delta -\frac{(1+z)}{{H}} (n^i\partial_i)^2 V ~~\mbox{where}~~ v_i=\partial_i V.
\ee
Here $v^i$ is the galaxy peculiar velocity and $n^i$ is the direction of the galaxy, and we use the Newtonian gauge. The Newtonian-Kaiser approach needs to be corrected at the theoretical level by including both types of relativistic effects.

\subsection{Nonlinear relativistic primordial correction}
An exactly Gaussian distribution of the primordial curvature perturbation translates into an exactly Gaussian distribution of density perturbations in the Newtonian approximation, where the Poisson equation is $\nabla^2\Phi=4\pi G a^2\rho\delta$ at all perturbative orders. In GR, the Newtonian Poisson equation is not correct at second order---there is a relativistic nonlinear correction in the GR  constraint equation that reduces to the Poisson equation at first order. This constraint links $\Phi$ to $\delta$ in the primordial Universe. Consequently, an exactly Gaussian distribution of primordial curvature perturbations does {\em not} lead to a Gaussian distribution of density perturbations, even on super-Hubble scales. We emphasise that this is a primordial correction and not a result of nonlinear evolution.

The effective local PNG parameter that describes this primordial GR correction on large scales (beyond the equality scale) is  \citep{Bartolo:2005xa,Verde:2009hy} 
\be \label{fgr}
\fng \simeq -2.2 ~~\mbox{(LSS convention)}.
\ee
(See also \citealt{Hidalgo:2013mba,Bruni:2013qta,Villa:2014foa,Camera:2014bwa}.)
The appropriate fiducial value for a concordance model is therefore not $\fnl=0$ but $\fnl\simeq-2.2$.
For large-scale structure, the best-fit value of $\fnl$  must be corrected as in \eqref{fcor}.
Note that this correction does not apply to PNG in the CMB, which is independent of the Poisson constraint. There are other nonlinear GR effects in the CMB which are accounted for in the \textit{Planck} constraint \eqref{fpg} \citep[][]{Ade:2015ava}.

\subsection{Linear relativistic lightcone effects}

The Kaiser RSD term in \eqref{dk} is the dominant term on sub-Hubble scales and at low redshifts of a more complicated set of first-order relativistic terms that arise from observing along lightrays which traverse the intervening large-scale structure. 

The first relativistic term is the lensing convergence, 
\be
\kappa =\int_0^\chi d\tilde\chi\,(\chi-\tilde\chi){\tilde\chi \over\chi}\nabla_\perp^2 \varphi,
\ee
which can make a significant contribution at higher redshifts on sub-Hubble scales. 
Here $\chi$ is the line-of-sight comoving distance   and $\nabla_\perp^2$ is the Laplacian on the screen space. Lensing affects the observed number density in two competing ways---enhancing it by bringing faint galaxies into the observed patch, and reducing it by broadening the area of the patch. The competition is mediated by the magnification bias, 
\begin{equation}
\mathcal Q=-\frac{\partial\ln N_{g}}{\partial\ln\mathcal F}\Big|_{\mathcal \mathcal F_\ast},\label{mbq}
\end{equation}
where $N_{g}(z,\mathcal F>\mathcal F_{\ast})$ is the background galaxy number density at redshift $z$ and with flux above the  survey limit $\mathcal F_{\ast}$. Therefore, the contribution of lensing to \eqref{dk} is $2(\mb-1)\kappa$. 

The remaining relativistic contributions are local and integrated terms, 
\be
\delta^{\rm obs}_g =\delta^z_g
+2\left(\mathcal Q-1\right) \kappa + \delta_{\rm loc}+\delta_{\rm  int}.
 \label{delgr}
\ee
In $\delta^z_g$ we use the comoving-synchronous overdensity, 
\be
\dco=\delta -{3H\over (1+z)}V,
\ee
in order to define the bias consistently on horizon scales \citep{Challinor:2011bk,Bruni:2011ta,Jeong:2011as}. Note that the first-order GR Poisson equation in Newtonian gauge is $\nabla^2\varphi=4\pi G a^2\rho\dco$. 
The additional relativistic terms in \eqref{delgr} are \citep{Yoo:2009au,Yoo:2010ni,Bonvin:2011bg,Challinor:2011bk,Jeong:2011as,Bertacca:2012tp,Jeong:2011as}
\begin{align}
\delta_{\rm loc}&=  {\left(3-b_e\right)H\over(1+z)} V + An^i \partial_i V +(2\mb-2-A)\varphi + {\dot\varphi \over {H}}, \nonumber\\
\delta_{\rm  int}&= 4  \frac{\left(1-\mb\right) }{\chi}\int_0^{\chi} {d\tilde \chi}\,\varphi  - 2A \int_0^{\chi} {d\tilde \chi} {\dot\varphi \over (1+z)}.
\label{delgr2}
\end{align}
Here, $b_e$ is the evolution bias, which reads
\be\label{eb}
b_e=-{\partial\ln (1+z)^{-3}N_g\over\partial\ln (1+z)},
\ee
and the factor $A$ is
\be
A=b_e -2\mb-1-\frac{\dot{H}}{{H}^2}+\frac{2\left(\mb-1\right)(1+z)}{\chi {H}}.
\ee

The local term $\delta_{\rm loc}$ has Doppler and SW type contributions. The integrated term $\delta_{\rm  int}$ contains time-delay and ISW contributions. These relativistic terms can become significant near and beyond the Hubble scale \citep{Yoo:2009au,Yoo:2010ni,Bonvin:2011bg,Challinor:2011bk,Jeong:2011as,
Bertacca:2012tp,Yoo:2012se,Raccanelli:2013dza,DiDio:2013bqa}. 

The growth of relativistic effects occurs on the same scales where the effect of PNG is growing through the galaxy bias of \eqref{eq:bias-NG}. Consequently,  relativistic lightcone effects can be confused with the PNG contribution \citep{Bruni:2011ta,Jeong:2011as,Yoo:2012se,Raccanelli:2013dza,Camera:2014bwa}. In order to remove this theoretical systematic when probing ultra-large scales, it is necessary to include all the relativistic lightcone effects in an analysis of PNG in galaxy surveys \citep{Maartens:2012rh,Camera:2014bwa}.

\section{Bias on $\fnl$ induced by disregarding GR effects}
If the nonlinear GR correction given by \eqref{fcor} is ignored, there will be an obvious and immediate bias  of $2.2$ in the best-fit $\fnl$, independent of the galaxy survey properties. We now show that ignoring the linear GR lightcone effects in \eqref{delgr} produces a further bias in the best-fit $\fnl$, which depends on the galaxy survey properties.

As a reference experiment, we use an SKA-like galaxy redshift survey. The number counts $N_{g}(z,\mathcal F>\mathcal F_{\ast})$ have been computed from simulations \citep{Camera:2014bwa}, thus avoiding unphysical assumptions on key survey parameters like magnification bias. In the case of \textit{Euclid}, the magnification bias is not available. But our main results are not particular to the SKA and are expected to apply qualitatively also to \textit{Euclid}-like surveys.

We use the angular power spectrum, $C_\ell$, where
\be
\langle\delta^{\rm obs}_g({\bf n},z) \delta^{\rm obs}_g({\bf n}',z') \rangle
=\sum_\ell{(2\ell+1)\over 4\pi}C_\ell(z,z') {\cal L}_\ell({\bf n} \cdot {\bf n'}),
\ee
and ${\cal L}_\ell$ are Legendre polynomials. The $C_\ell$ are computed using \textsc{camb}\_sources \citep{Challinor:2011bk}. The Fisher matrix formalism is widely used in parameter estimation and survey design. The formalism may also be employed to evaluate the systematic bias in the best-fit values of a cosmological parameter set, $\vartheta_\alpha$, arising from the incorrect treatment of correlations in the theoretical template---for example, in the case when we disregard GR effects.

We split the source distribution into redshift bins  to construct a tomographic matrix $\mathbf C_{\ell}$ whose entries $C^{ij}_\ell$ are the angular power spectra for the $(i,j)$ bin pair. 
Assuming a Gaussian likelihood function for the model parameters, the bias on the best-fit $\fnl$ value, $b(\fnl)$, can be estimated as \citep[see e.g.][]{Kitching:2008eq}
\begin{multline}
b(\fnl)=f_{\rm sky}\sum_{\alpha,\ell}\frac{2\ell+1}{2}\big(\tilde{\mathbf F}^{-1}\big)_{\fnl\alpha}\\\times\mathrm{Tr}\Big[\big(\tilde{\mathbf C}_{\ell}+\mathbf{\mathcal N}_\ell\big)^{-1}\frac{\partial\tilde{\mathbf C}_{\ell}}{\partial\vartheta_\alpha}\big(\tilde{\mathbf C}_{\ell}+\mathbf{\mathcal N}_\ell\big)^{-1}\big(\mathbf C_{\ell}-\tilde{\mathbf C}_{\ell}\big)\Big],
\end{multline}
where $ f_{\rm sky}$ is the fraction of the celestial sphere surveyed.
The absence or presence of a tilde denotes the case with and without GR corrections, respectively. $(\tilde{\mathbf F}^{-1})_{\fnl\alpha}$ is a row of the inverse of the Fisher matrix (without GR corrections), which includes the auto- and cross-terms between $\fnl$ and all the $\vartheta_\alpha$.  The experimental noise on a measurement of $\mathbf C_{\ell}$ is   ${\mathcal N}^{ij}_\ell=\delta^{ij}/\bar N_g^i$ where $\bar N_g^i$ is the number of galaxies per steradian in the $i$-bin.

For the SKA-like survey we adopt to estimate the impact of the bias on $\fnl$, we consider a flux rms of  $3\,\mu$Jy \citep[for detailed specifications, see ][]{Camera:2014bwa}. As noted in \citet{Camera:2014bwa},  PNG effects increase as redshift increases, but this is countered by an increase in the noise with $z$. As a result, an intermediate redshift interval has a more optimal balance between PNG effects and noise, thus yielding the tightest constraints on $\fnl$. We choose a redshift range $0.5\leq z\leq2.5$, subdivided into 20 bins of constant width $\Delta z=0.1$. 

\begin{figure*}
\centering
\includegraphics[width=\columnwidth]{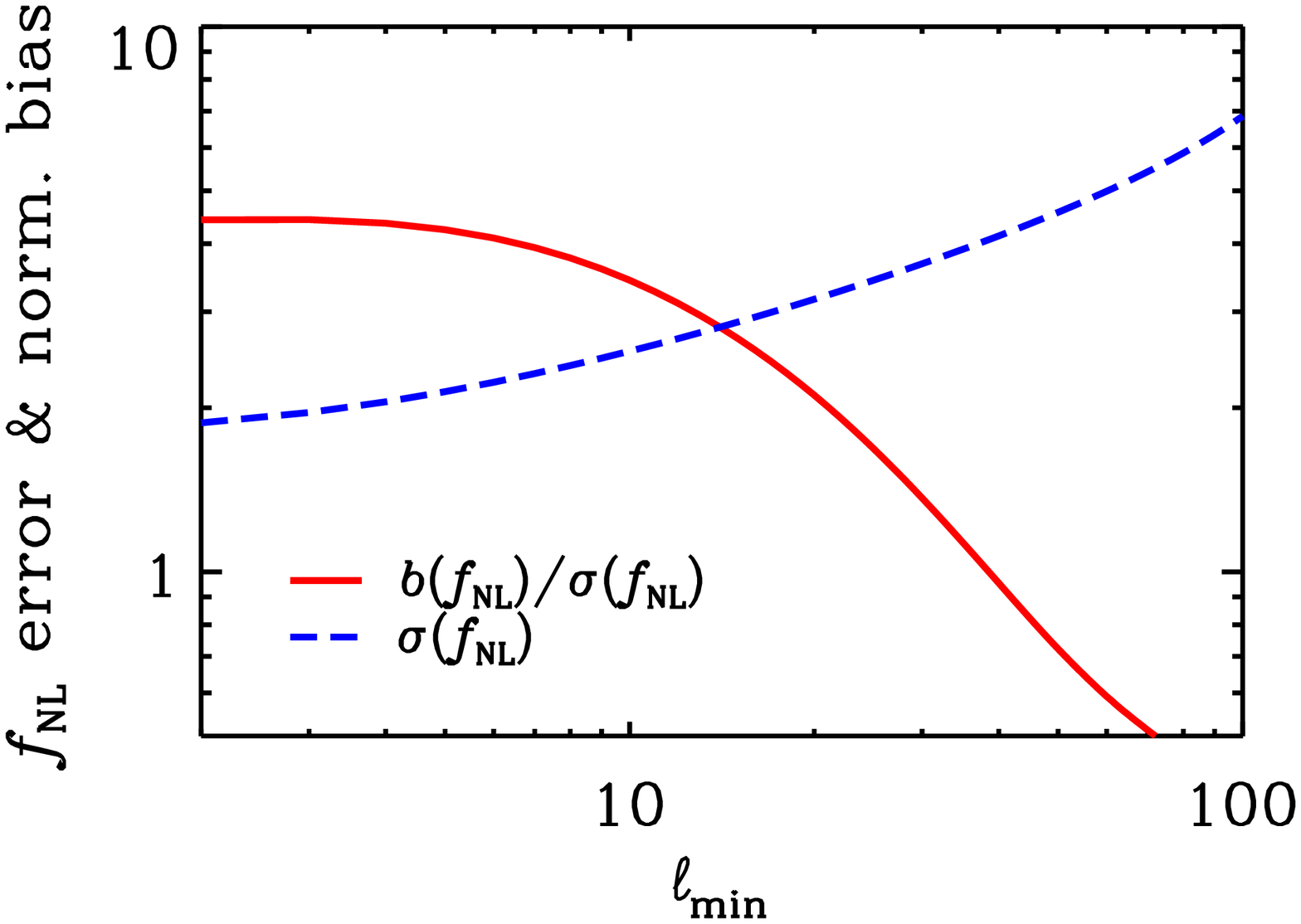}
\includegraphics[width=\columnwidth]{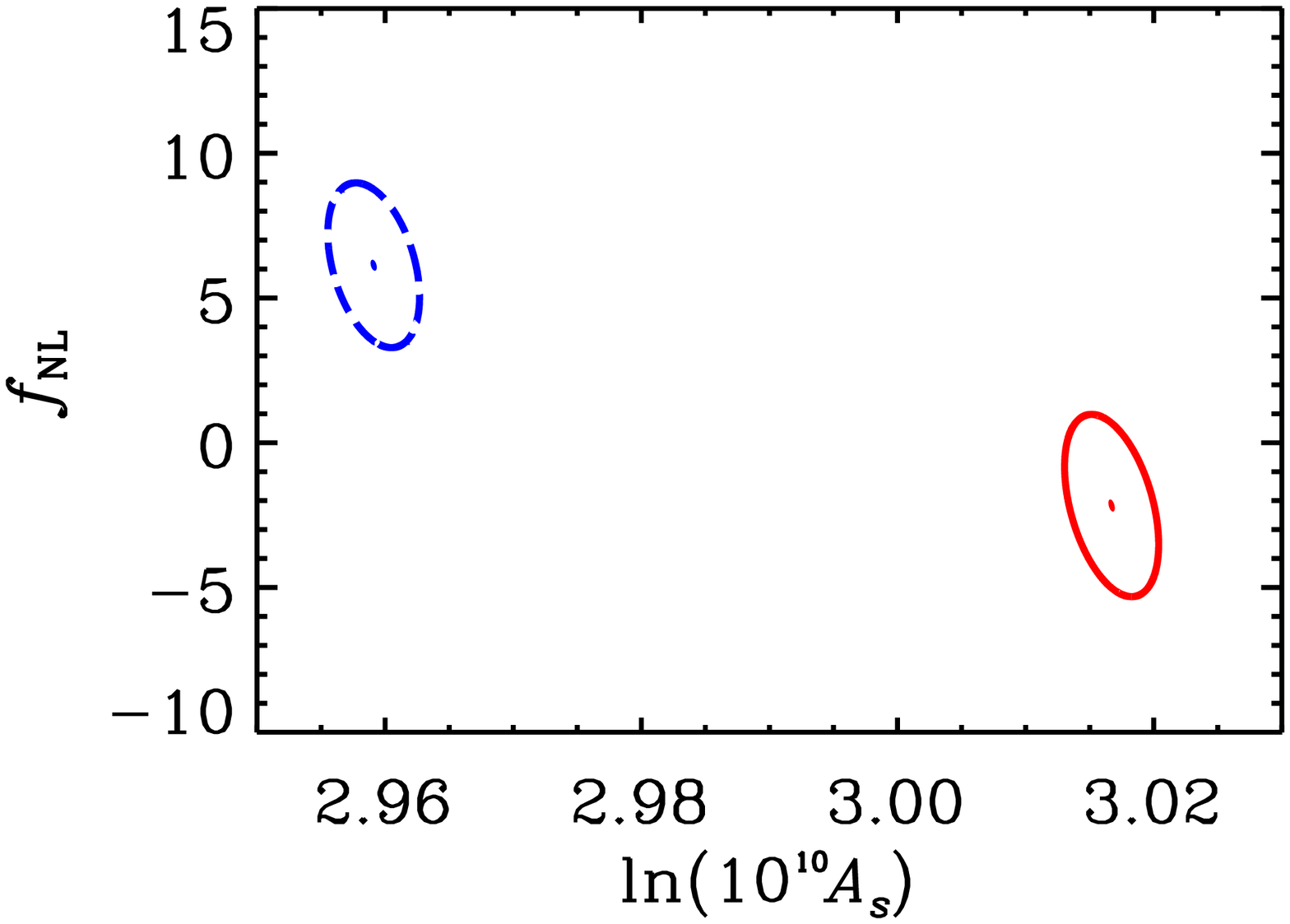}
\caption{{\em Left:} Normalised bias on the best-fit $\fnl$ value (solid, red curve) as a function of the minimum angular multipole, $\ell_{\rm min}$.
{\em Right:} 1$\sigma$ joint contours in the $(\fnl,A_s)$ plane, showing the shift in the best-fit (dots) from neglecting GR lightcone effects. The solid (red) ellipse includes all GR effects, while they are neglected for the dashed (blue) contour. Here we fix $\ell_{\rm min}=2$.
}\label{fig:DfNLvsLmin}
\end{figure*}

The fiducial model is a concordance model with \textit{Planck} best-fit parameters \citep{Ade:2013zuv}. The concordance model has Gaussian primordial fluctuations, and we incorporate the nonlinear GR correction by using a fiducial value 
\be
f_{\rm NL}^{\rm fid}\simeq-2.2.
\ee
The forecast marginal error $\sigma(\fnl)$ is shown in Fig.~\ref{fig:DfNLvsLmin} (left panel) as a function of the maximum angular scale, i.e. minimum angular multipole $\ell_{\rm min}$. Also shown is the induced bias on $\fnl$ in units of $\sigma(\fnl)$ when GR lightcone effects are disregarded. The general trend of the two curves follows from the basic properties of PNG and GR lightcone effects: the smaller $\ell_{\rm min}$ is (i.e.\ the larger the maximum angular scale probed by the survey), the stronger is the signal of PNG and of the GR effects. Therefore $\sigma(\fnl)$ is smaller, since we have more information from the scales where PNG is stronger, while the bias on $\fnl$ induced by neglecting GR lightcone effects is larger.

Figure~\ref{fig:DfNLvsLmin} (right panel) illustrates how the best-fit values and 1$\sigma$ contours move in the $(\fnl,A_s)$ plane, where $A_s$ gives the primordial amplitude of the curvature perturbation.  The solid, red contour depicts the forecast 1$\sigma$ two-parameter error contour that would be obtained if we consistently accounted for GR lightcone effects in the analysis. The dashed, blue ellipse refers to the case where we neglect these GR corrections, thus biasing the best-fit value of our measurement. Note that the marginal errors themselves change only slightly when GR lightcone effects are neglected. 

To be more conservative, we include $A_s$ in the analysis as it is the cosmological parameter that is most degenerate with $\fnl$ on large scales;  $\fnl$ is known not to be strongly degenerate with the other, standard cosmological parameters---particularly on the extremely large scales of interest here.

Our main concern here is how to extract PNG from the galaxy power spectrum---i.e.\ how to deal with the induced bias on $\fnl$. The even greater bias on $A_s$ in Fig.~\ref{fig:DfNLvsLmin} should not be taken at face value. The magnitude of $b(A_s)$   is mainly due to the smallness of $\sigma(A_s)$.  Hence, whilst $\sigma(\fnl)$ is reasonably accurate, the marginal error on $A_s$ is not to be regarded as an actual parameter forecast, as it can be further pinned down through measurements on smaller scales. Its role, here, is more that of a nuisance parameter.

\section{Conclusions}
There are two types of relativistic effects that correct the standard analysis of local PNG in the galaxy power spectrum and both are essential in order to avoid a bias on the best-fit value of $\fnl$. A nonlinear GR correction to the primordial Poisson equation leads to a shift in the best-fit $\fnl$, given by \eqref{fcor}, which is independent of galaxy survey properties. In addition, there are survey-dependent corrections due to linear GR effects from observing number counts on the past lightcone---as given in \eqref{delgr}-\eqref{delgr2}.

The behaviour in Fig.~\ref{fig:DfNLvsLmin}  (left) of the forecast error and the normalised bias on $\fnl$  follows since both PNG and GR lightcone effects grow as the scales probed by the survey increase, i.e.\ as $\ell_{\rm min}$ decreases. If we probe the clustering properties of  cosmic structure on extremely large scales in order to reduce $\sigma(\fnl)$, but we ignore the GR corrections, then we pay the price of a serious bias on the measurement of the best-fit $\fnl$. 

In the case of the reference survey used here, this theoretical bias  shifts the best-fit value of $\fnl$ by $\gtrsim 4\sigma$ in the optimal multipole range $2\lesssim\ell_{\rm min}\lesssim8$ (Fig.~\ref{fig:DfNLvsLmin}, right). In this range the  normalised bias on $\fnl$ is almost constant, indicating that the bias itself is proportional to the error, i.e. $b(\fnl)\propto \sigma(\fnl)$ for $\ell\lesssim 8$. The bias on $\fnl$ is above 2$\sigma$ for $\ell_{\rm min}\lesssim 20$.

For current surveys, which are unable to probe such large angular scales, only the nonlinear correction \eqref{fcor} is relevant for avoiding bias in the best-fit $\fnl$.  The bias from neglecting GR lightcone effects is currently negligible. However, future surveys, such as \textit{Euclid} and the SKA, will dramatically decrease the error $\sigma(\fnl)$ by including very large scales, and in this case the bias on the best-fit $\fnl$ from ignoring GR effects would be significant. In the example of Fig.~\ref{fig:DfNLvsLmin}  (right), the bias would in particular lead us to falsely conclude that the primordial Universe was non-Gaussian. This is important also because an incorrect treatment of PNG may lead to inaccurate reconstructions of other, standard cosmological parameters \citep{Camera:2014dia}.

The bias on the best-fit $\fnl$ is not special to the fiducial best-fit value of $\fnl\simeq-2.2$. We have checked that for $|\fnl|\lesssim10$, consistent with the \textit{Planck} constraint \eqref{fpg}, $b(\fnl)$ does not change appreciably. There is however a level of sensitivity to the physical survey features, i.e. $b, \mb$ and $b_e$. We have been careful not to arbitrarily choose these  functions, but to derive them from simulations for a proposed experiment, the SKA. We do not expect that the results would change significantly for a \textit{Euclid}-like spectroscopic survey. 

Can we by-pass the problem by using the multi-tracer method \citep{Seljak:2008xr,Ferramacho:2014pua}? No---the multi-tracer method is accessing precisely the largest scales where the GR effects are strongest, and hence the same bias problem will emerge if these effects are ignored.


\subsection*{Acknowledgments}
SC acknowledges support from the European Research Council under the EC FP7 Grant No. 280127 and FCT-Portugal under Post-Doctoral Grant No. SFRH/BPD/80274/2011. MS and RM  are supported by the South African Square Kilometre Array Project and the South African National Research Foundation. RM acknowledges support from the UK Science \& Technology Facilities Council (grant ST/K0090X/1). 

\bibliographystyle{mn2e}
\bibliography{../../../../Bibliography}

\bsp

\label{lastpage}

\end{document}